\documentclass[
 reprint,
superscriptaddress,
amsmath,amssymb,
aps,
prr
]{revtex4-2}

\usepackage{graphicx}
\usepackage{bm}
\usepackage{hyperref}
\usepackage{enumitem}
\usepackage{xcolor}
\definecolor{darkpowderblue}{rgb}{0., 0.2, 0.6}
\usepackage{dutchcal}
\usepackage{orcidlink}

\begin{document}

\title{Measurement-induced transitions beyond Gaussianity:\\ a single particle description}

\author{Luca Lumia~\orcidlink{https://orcid.org/0000-0002-7783-9184}}
\affiliation{International School for Advanced Studies (SISSA), Via Bonomea 265, I-34136 Trieste, Italy}

\author{Emanuele Tirrito}
\affiliation{The Abdus Salam International Centre for Theoretical Physics (ICTP), Strada Costiera 11, 34151 Trieste, Italy}
\affiliation{Pitaevskii BEC Center, CNR-INO and Dipartimento di Fisica,
Università di Trento, Via Sommarive 14, Trento, I-38123, Italy}

\author{Rosario Fazio~\orcidlink{0000-0002-7793-179X}}
\affiliation{The Abdus Salam International Centre for Theoretical Physics (ICTP), Strada Costiera 11, 34151 Trieste, Italy}
\affiliation{Dipartimento di Fisica, Università di Napoli “Federico II”, I-80126 Napoli, Italy}

\author{Mario Collura}
\affiliation{SISSA, Via Bonomea 265, I-34136 Trieste, Italy}
\affiliation{INFN Sezione di Trieste, via Bonomea 265, I-34136 Trieste, Italy}


\begin{abstract}
	Repeated measurements can induce entanglement phase transitions in the dynamics of quantum systems. Interacting models, both chaotic and integrable, generically show a stable volume-law entangled phase at low measurement rates which disappears for free, Gaussian fermions. Interactions break the Gaussianity of a dynamical map in its unitary part, but non-Gaussianity can be introduced through measurements as well. By comparing the entanglement and non-Gaussianity structure of different protocols, we propose a new single-particle indicator of the measurement-induced phase transition and we use it to argue in favour of the stability of the transition when non-Gaussianity is purely provided by measurements.

\end{abstract}
\maketitle

\begin{figure*}[t]
\centering\includegraphics[width=0.9\textwidth]{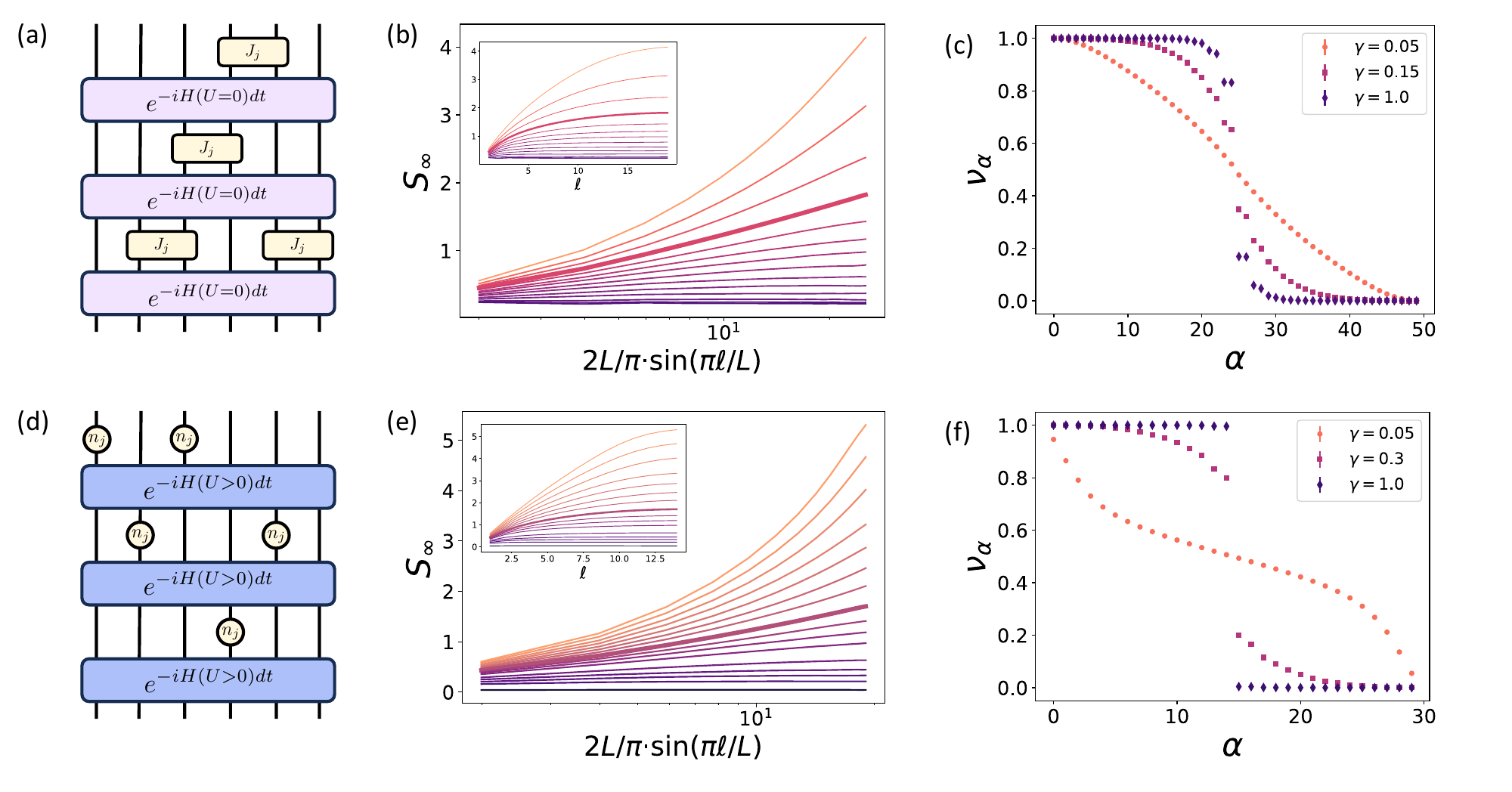}
\caption{(a) Graphical representation of the monitored dynamics associated to hopping fermions with measurements of the current; the first line always refers to this case, while the second line refers to the trajectories generated by the interacting Hamiltonian with $n_j$ measurements represented in (d). (b),(e) Stationary entanglement entropies at $\gamma\in[0.05,\,2.0]$ for different bipartitions $\{1,2,...,\ell\}$ on a chain with fixed length ($L=40$ for $J_j$, $L=30$ for $n_j$). The volume-law growth of $S_\infty$ in the interacting case appears in (e) sufficiently far from the boundary where it is affected by the finite size. At the critical point the entanglement is scale invariant and shows a linear growth w.r.t. the logarithm of the chord length $x=2L/\pi\sin(\pi\ell/L)$ as shown in (b),(e) (notice the log scale in the x axis). (d),(h) Occupations of the natural orbitals at different measurement rates: volume-law steady states are associated to a smooth spectrum, while in the area-law phase the spectrum is gapped.} 
\label{fig: main}
\end{figure*}

\section{Introduction}

Entanglement is a fundamental feature of quantum mechanics and in the last decades it has become a resource of primary importance for quantum information tasks \cite{Horodecki2009}, as well as a key tool to describe the physics of quantum many-body systems \cite{Amico2008}, both in and out of equilibrium. Typical eigenstates in the middle of the energy spectrum  have extensive entanglement entropy and this plays an important role in the understanding of thermalization and of its exceptions \cite{Deutsch1991, Srednicki1994, Polkovnikov2011}. In contrast, ground states of gapped Hamiltonians are short-range entangled, because local interactions generate quantum correlations only between sufficiently near degrees of freedom, correspondingly producing entanglement entropies proportional to the area of the boundary between the subsystems\cite{Hastings2007, Wolf2008}. A notable exeption is provided by 1D critical points, where the divergence of the correlation length is associated to a universal logarithmic scaling of the entanglement \cite{Calabrese2004}. 

Phase transitions typically arise as a result of competing interactions or driving mechanisms that steer a many-body system towards macroscopically different states. An interesting competition in quantum systems is found in monitored dynamics \cite{Wiseman2009}, where measurements contrast the entanglement generation induced by the unitary evolution. When the measurement rate is increased, this competition is able to drive a dynamical transition called \emph{measurement-induced phase transition} (MIPT), that was first witnessed in simulated hybrid quantum circuits \cite{Li2018,Skinner2019,Li2019,Szynieszewski2019,Choi2019,Gullans2020,Zabalo2020,Bao2020,Jian2020,Ippoliti2021,Block2022} and some signatures of the phase transition have now been experimentally observed \cite{Noel2022,Koh2023,Google2023}. Following the dynamics of a pure state, the quantum trajectories generated by measurements reach a steady state that undergoes a transition from a volume-law entangled phase to an area-law entangled phase characterized by the Zeno effect. 
The stability of extensive entanglement for small measurement rates can be understood as a consequence of scrambling: a unitary evolution spreads any information that was initially encoded in localized degrees of freedom, protecting it from local projective measuremtents because fully retrieving it would require measuring global operators \cite{Choi2019,Li2021}. 
When the unitary evolution is generated by a generic non-integrable Hamiltonian, it is natural to expect the same phenomenology of random circuits since they are a model of chaotic dynamics. This was indeed observed and, notably, even at integrable points different properties in the MIPT have not been identified \cite{Fuji2020,Tang2020,Xing2023,Cecile2024}.
%
%
Other entanglement phases can be introduced by long range models \cite{Block2022}, but restricting to short-range systems an important exception is provided by free fermions with local occupation measurements: the volume-law is much more unstable and disappears at any finite rate of measurements \cite{Cao2019,Alberton2021,Buchhold2021, Turkeshi2021,Coppola2022,Turkeshi2022,Kells2023,Fava2023,Poboiko2023,Chahine2023,Jian2023}, leaving the place to a subextensive region whose nature is currently debated. This regime was initially associated to a logarithmic BKT-like critical phase \cite{Alberton2021, Buchhold2021}, while more recent numerical works and theoretical studies based on a mapping to non-linear sigma models argue that the logarithm should saturate to area-law in the thermodynamic limit \cite{Coppola2022,Poboiko2023}. Different symmetries may play a role in determining the properties and the stability of the transition \cite{Loio2023,Fava2023}.

An important difference between free fermions and other integrable models is Gaussianity \cite{Peschel2009}, which is preserved by measurements of particle numbers. The volume law-entangled phase arises with interactions, but a similar Gaussianity-breaking can be achieved with suitable measurements. Is the non-Gaussianity introduced by measurements sufficient to stabilize the phase transition? In this work, we address the question by comparing the entanglement and the non-Gaussianity structure of hopping fermions with repeated measurements of their current, and we find that non-Gaussian measurements are able to restore the volume law/area law MIPT. A relevant role is played by the one-body reduced density matrix of the evolving state, which encodes all two-point correlations and allows to measure non-Gaussianity. Its eigenstates, called natural orbitals, are single-particle states adapted to the many-body problem and we find that the information about the original state that they retain is sufficient to witness the transition, as their occupation spectrum shows the opening of a gap in the area-law phase. This provides a new indicator of the MIPT, which is remarkably simple to access since it requires only the calculation of two-point functions.

\vspace{3pt}

\section{Model and dynamical protocols}
We study the quench dynamics of a system evolving under the combined effect of a hermitian Hamiltonian $H$ and repeated projective measurements of an observable ${Q}$ at a finite rate $\gamma$. Consider an open chain with $L$ sites and hopping spinless fermions with Hamiltonian
\begin{equation}
\label{eq: hamiltonian}
    H = -\frac{1}{2} \sum_{j=1}^{L-1} \left(c^\dagger_j c_{j+1} + c^\dagger_{j+1}c_j\right) + U \sum_{j=1}^{L-1} n_jn_{j+1}\,,
\end{equation}
where everything is expressed in units of the hopping energy.
Having defined $N = \sum_j n_j$, $[H,N]=0$ and the total number of fermions is a conserved $U(1)$ charge. The Heisenberg equation of motion of its local density $n_j$ takes the form of a continuity equation that defines the current $J_j = -{i}/{2} \,(c^\dagger_j c_{j+1} - c_{j+1}^\dagger c_j)$.

\vspace{1pt}

\subsection{Monitored evolution}

The system is initially prepared in a product state $|\psi_0\rangle$ and if no measurements take place it evolves as $|\psi(t)\rangle=\exp(-iHt)|\psi_0\rangle$.  In every time interval of width $dt$ there is a probability of $\gamma dt$ of measuring a local operator ${Q}_j$ and as a result the state collapses into the eigenspace associated to one of its eigenvalues $q$, chosen with probability given by Born's rule $p_\psi(q)=\langle \psi | \Pi^q_j |\psi\rangle$, where $\Pi^q_j$ is the projector on the $q$-th eigenspace. A representation of the two protocols is given in Fig.~\ref{fig: main}. 
Measurements are local, but not necessarirly single-site: we will consider occupation numbers ${Q}_j=n_j$ and currents ${Q}_j=J_j$, that \mbox{are bond} operators. In both cases $[{Q}_j,N]=0$ and the $U(1)$ symmetry is preserved, but notice that neighbouring currents do not commute and the order in which their measurements are performed matters. 
Current operators have three eigenvalues $J=\pm1/2,\,0$, \mbox{associated} to the eigenspaces \mbox{$\mathcal{V}_{\pm1/2}=\text{span}\{
(c^\dagger_{j+1}\pm ic_j^\dagger)|00\rangle\}$} and $\mathcal{V}_0 = \text{span}\{|00\rangle, \,c^\dagger_j c^\dagger_{j+1}|00\rangle\}$ on the two-site vacuum $|00\rangle$. 
The eigenstates in $\mathcal{V}_{\pm1/2}$ are Gaussian, while the last eigenspace is degenerate and we expect to introduce non-Gaussianity everytime we project on it, because linear combinations of Gaussian states do not generally preserve their Gaussianity.
%
%
%
%
Following its evolution conditioned on measurement outcomes, the state remains pure and evolves along a stochastic trajectory called \emph{quantum trajectory} \cite{Wiseman2009}, defined as a realization of the process that solves the stochastic Schrödinger equation (SSE)
\begin{equation}
\begin{split}
\label{eq: SSE}
    d|\psi(t)\rangle & =  \Biggl\{dt\Biggl[-iH + \frac{\gamma}{2}
        \sum_{j,q}\left(\frac{\langle \Pi^q_j\rangle_t}{2} - \frac{\Pi_j^q}{2}\right) \Biggr] \\
    & + \sum_{j,q} dN_{jq}(t) \Biggl(\frac{\Pi_j^q}{\sqrt{\langle \Pi_j^q\rangle_t}} -1 \Biggr)\Biggr\}\,|\psi(t)\rangle\,.
\end{split}
\end{equation}
$dN_{jq}(t)=0,1$ s.t. $dN_{jq}(t)dN_{j'q'}(t) = \delta_{qq'}\delta_{jj'}dN_{jq}(t)$ and $\overline{dN_{jq}(t)} = \gamma dt\,\langle \Pi_j^q\rangle_t,\,\langle \Pi_j^q\rangle_t\equiv\langle\psi(t)|\Pi_j^q|\psi(t)\rangle\,$ are increments of independent Poisson variables that count the occurrences of each measurement \mbox{outcome} on the realization.  Each trajectory reaches a steady state and all quantities calculated over it \mbox{are stochastic.} Linear observables $\mathcal{O}$ do not provide further information about the steady state, because in that case $\overline{\text{tr}(|\psi(t)\rangle\langle\psi(t)|\,\mathcal{O})}=
\text{tr}(\rho\, \mathcal{O})$ and the mean state relaxes towards $\rho_\infty\propto\mathbb{I}$. For a discussion of the SSE and of the unconditional dynamics of $\rho$ see App. \ref{app: Conditional and unconditional dynamics}. The only way to observe the MIPT is to evaluate first a non-linear functional on the steady state of each trajectory $|\psi_{\infty}\rangle$ before taking the average over trajectories and not viceversa.
%
%
The most common indicator of MIPTs is the entanglement \mbox{entropy $S_A(\psi) = -\text{tr}(\rho_A\log{\rho_A})$, where $\rho_A = \text{tr}_{\bar{A}}(|\psi\rangle\langle\psi|)$} is the reduced density matrix of a subsystem $A$. Other non-linear quantities that witness the transition have been discussed \cite{Tirrito2023,Paviglianiti2023}, and here we describe how the transition affects the non-Gaussianity of the states. In particular, we find that a gap in the occupation spectrum of the natural orbitals opens up at the MIPT, as shown in Fig.~\ref{fig: main} and discussed later in more detail.

\subsection{Numerical simulation}

The measurement terms of the SSE associated to occupation measurements are quadratic and preserve the Gaussianity of a trajectory. 
%
We are interested in generating non-Gaussianity, either by interactions in the unitary part $-iH$ or by non-quadratic measurement terms like those provided by the eigenprojectors of the current (see App. \ref{app: Details on the simulation}), and to simulate such trajectories we resort to tensor network methods. By formulating the quantum channel in its Kraus representation, as shown in App.\ref{app: Conditional and unconditional dynamics}, the solutions of the SSE \eqref{eq: SSE} can be directly formulated as a tensor network of the kind represented in Fig. \ref{fig: main}.
The growth of entanglement is tamed by the measurements, enabling us to simulate the dynamics up to long times. Fermionic models can be implemented as a tensor network thanks to the Jordan-Wigner transformation, which maps their anticommuting degrees of freedom into qubits. The Hamiltonian \eqref{eq: hamiltonian} becomes an XXZ chain with longitudinal field, particle numbers $n_j$ become Pauli $\sigma^z_j$ operators and $J_j$ translates into the spin current. For more details on the simulations, see App. \ref{app: Details on the simulation}.

\section{Results}

We start by considering the quantum trajectories generated by evolving the Néel state on a chain with $L$ sites under the free Hamiltonian \eqref{eq: hamiltonian} with $U=0$ and random projective measurements of the current at a fixed rate $\gamma$. The evolution is simulated with a TEBD-like algorithm interspersed with measurements. To characterize the steady state we measure the entanglement entropies $S_A(t),\,A=\{1,...,\ell\}$ over left-right bipartitions and the correlation matrices $\langle c^\dagger_i c_j\rangle$. The observables are calculated with a late-time average on a each trajectory, for all times $t>t_{rel}$ after relaxation, together with the average over trajectories.

\subsection{Steady-state entanglement}

The stationary entanglement entropies, denoted by $ S_\infty(\ell)\equiv\overline{S_A(\infty)}$ for $|A|=\ell$, are shown in Fig. \ref{fig: main} from $\gamma=0.05$ to $\gamma=2.0$ and we check their scalings with both the subsystem size $\ell$ and the total size $L$. For frequent measurements the entanglement is upper bounded and the system is in the area law, while in the rare-measurement regime no clear volume law appears. 
At low measurement rates $0.05<\gamma<0.11$ we always see sub-extensive entanglement, that is initially super-logarithmic and appears to approach a $\log$ for sufficiently large subsystem sizes $\ell$. However, $\ell\approx L/2$ is also a region with strong finite-size effects and it is difficult to faithfully tell what the phase should be from the entanglement entropy alone.
%
%
%
To better understand the extent of finite-size effects, consider now the case of interacting Hamiltonian \eqref{eq: hamiltonian} with $U=1$ and Gaussian measurements of $n_j$, which is expected to exhibit a volume-law/area-law MIPT \cite{Fuji2020,Tang2020,Xing2023}. The results are presented again in Fig.~\ref{fig: main}. After the transient regime the stationary entanglement entropy ${S}_\infty(\ell)$ distinguishes two regimes with distinct scaling behaviours: for small measurement rates it is extensive ${S}_\infty(\ell)\sim\ell$, while for higher rates independent of the subsystem size ${S}_\infty(\ell)\sim \ell^0$. The difference can be appreciated for subsystems sufficiently far from the boundary, since the volume law shows an inflection for $\ell\approx L/2$ which is a finite-size effect and it is expected to grow unbounded for \mbox{$L\to\infty$.} In the thermodynamic limit, the two regimes correspond to stable dynamical phases separated by a MIPT, that for our limited sizes smoothens into a crossover. 
%
%
Notice how the finite-size effect close to half chain makes the growth of entanglement look similar to the current measurements case. Given our data, when the non-Gaussianity is induced by measurements it is difficult to discern if the thermodynamic rare-measurement behavior is supposed to be logarithmic, sub-extensive or volume-law. The entanglement barrier of MPS prevents us from reaching sufficiently small rates to unambiguously see past the crossover. In order to clarify the picture, we go beyond the entanglement classification by studying the non-Gaussianity structure of the stationary states. This will show the stability of the volume-law. 


%

\begin{figure}[t]
\centering
    \includegraphics[width=0.45\textwidth]{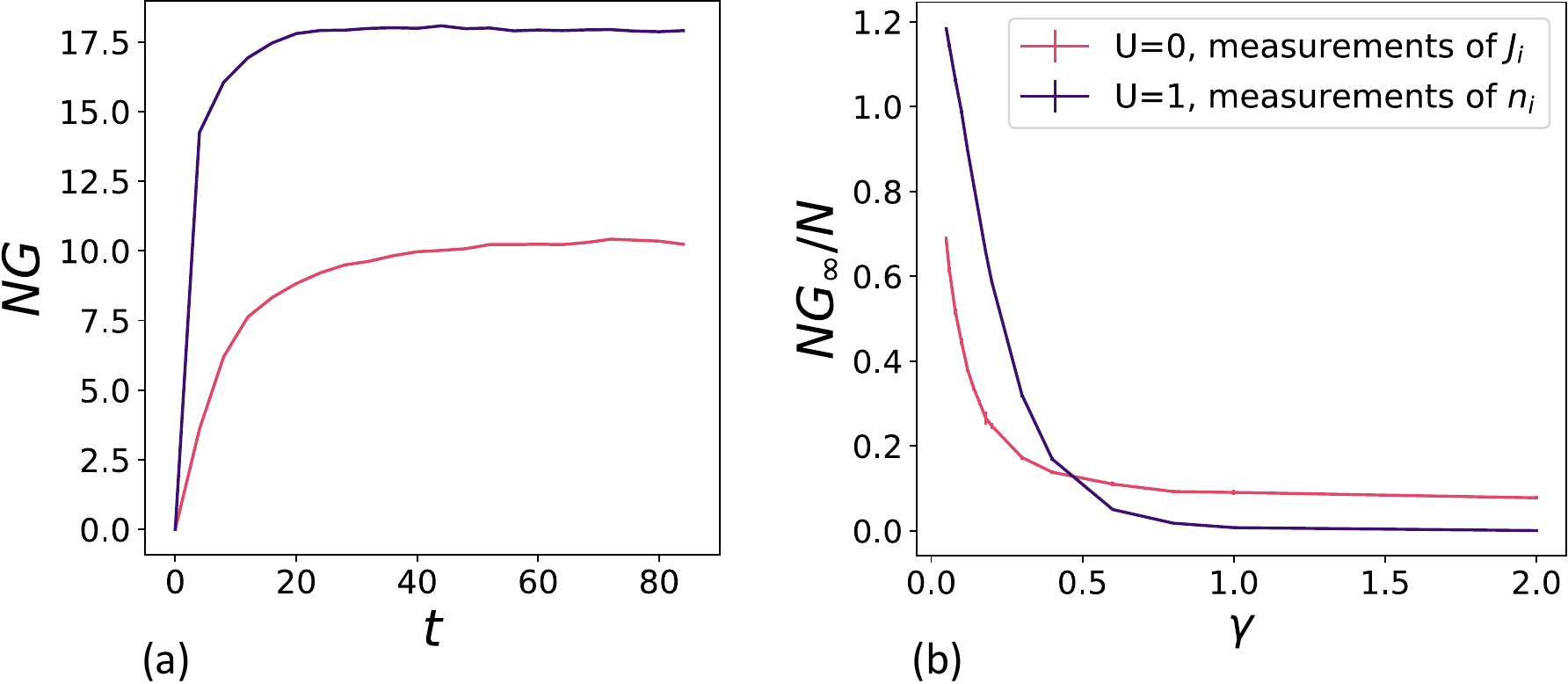}
\caption{Total non-Gaussianity for current measurements on a free Hamiltonian and for occupation measurements on an interacting Hamiltonian, (a) compares its time evolution for $\gamma=0.05$ in both cases and (b) the behaviour of its stationary value as a function of the measurement rate. Both graphs are generated for a chain with $L=30$ sites and the stationary values have been renormalized by the particle number because $NG$ is extensive. Gaussianity is immediately broken by interactions as well as current measurements, but for current measurements the relaxation process is slower and the overall amount of non-Gaussianity that they introduce is smaller.} 
\label{fig: NG}
\end{figure}

\subsection{Total non-Gaussianity}


Gaussian states can be identified by checking whether Wick's theorem holds, but it is simpler to look at the correlation matrix $C_{ij}(\psi)=\langle \psi | c^\dagger_i c_j | \psi \rangle$, which is the matrix representation of the one-body reduced density operator $\rho^{(1)}(\psi) = \sum_{ij} C_{ij}(\psi)\,c^\dagger_i c_j$. Its eigenvectors, the natural orbitals $|\phi_\alpha\rangle$ s.t. ${\rho}^{(1)}(\psi)|\phi_\alpha\rangle = \nu_\alpha|\phi_\alpha\rangle$, are a basis of single-particle states which retains information about the many-body correlations of the state. Our dynamics is always number-conserving, so there is no need to consider anomalous correlations $\sim cc,\,c^\dagger c^\dagger$. The eigenvalues $\nu_\alpha$ are occupation numbers and $N=\sum_{\alpha}\nu_\alpha$ is the total number of particles. Pure Gaussian states are Slater determinants; the normal modes coincide with the natural orbitals and the eigenvalues of $C_{ij}$ can only be $0$ or $1$ if the mode is respectively empty or occupied, so that $\nu_\alpha \neq0,1$ characterize the departure of a state from Gaussianity. A natural way to measure the non-Gaussianity of a pure state $|\psi\rangle$ is then to use a binary entropy
\begin{gather}
    NG(\psi) = \sum_\alpha H_2(\nu_\alpha)\,,\\
    H_2(\nu)= -\nu_\alpha\log\nu - (1-\nu_\alpha)\log{(1-\nu_\alpha)}\,.
\end{gather}

\noindent As shown in App.\ref{app: Total non-Gaussianity}, this definition is related to the distance between $|\psi\rangle$ and the closest Gaussian state. The properties of the quantum relative entropy grant its well-behavedness from the point of view of resource theory.
%
%

Let us consider first the evolution of  $NG(t)\equiv\overline{NG(\psi(t))}$ for the interacting Hamiltonian \eqref{eq: hamiltonian} with $U=1$ and Gaussian $n_j$ measurements and for the free Hamiltonian with $U=0$ and current measurements. The results are shown in Figure \ref{fig: NG} and as expected current measurements break Gaussianity. The profile has an initial transient regime where it increases in parallel to the amount of non-Gaussian operations performed, before saturating to a stationary value. 
%
%
Thanks to the continuous non-Gaussianity pumping the behaviour is similar in the two cases, even if the stationary regime is provided by different balances: in the former the $NG$ is brought by the unitary evolution and reduced by measurements, while in the latter measurements are what introduces $NG$. 

It is natural to ask whether the non-linear quantity $NG_\infty\equiv\overline{NG(\psi_\infty)}$ is able to detect the MIPT and in Fig. \ref{fig: NG} we shown that it is not: it is always smoothly decreasing as a function of the measurement rate $\gamma$. When we measure $n_j$ ${NG}_{\infty}(\gamma)\to0$ for $\gamma\to\infty$. This is a consequence of the “freezing” process typical of the Zeno effect: for $\gamma\to\infty$ $n_j$ is measured at all sites and the state is locked in the Néel configuration, which is Gaussian. 
%
%
In the opposite limit $\gamma\to0$ the non-Gaussianity increases, getting close to its maximal value $S_{G,max}=2N\log{2}$, corresponding to a flat spectrum $\nu_\alpha=1/2$ for all $\alpha=1...L$. This condition is compatible with the particle number conservation from the Néel state $N=\sum_\alpha\,\nu_\alpha=L/2$. 
For current measurements there is an obvious difference with $n_j$ measurements in the large rate limit: ${NG}_{\infty}(\gamma)$ has a finite asymptote for $\gamma\to\infty$. Current measurements do not commute and even when all sites are measured it is impossibile to project the state in a product state because of $J=0$ outcomes, causing a finite non-Gaussianity remnant. 
The functional shape is similar, with an initial non-Gaussianity peak and a monotonous decay. This behaviour was expected in the $n_j$ case, where the effect of measurements is to push the state closer to Gaussianity, but it is totally non trivial when measuring currents, because now measurements are what introduces non-Gaussianity in the dynamics as well. For $\gamma=0$ the state is Gaussian at all times and ${NG}_{\infty}=0$; introducing a small measurement rate one could expect an initial regime with increasing non-Gaussianity that is instead absent. As soon as a $\gamma>0$ is turned on ${NG}_{\infty}(\gamma)$ jumps to its maximal value, meaning that the two limits $\gamma\to0$ and $t\to\infty$ do not commute. Indeed a small absolute number of measurements introduces little non-Gaussianity, since for short times ${NG}(t)\ll1$, but here we are taking the stationary limit first and for every finite $\gamma$ the state has undergone an extensive amount of measurements which allowes it to reach its long-time non-Gaussianity balance.

\begin{figure}[t]
\centering
\includegraphics[width=0.48\textwidth]{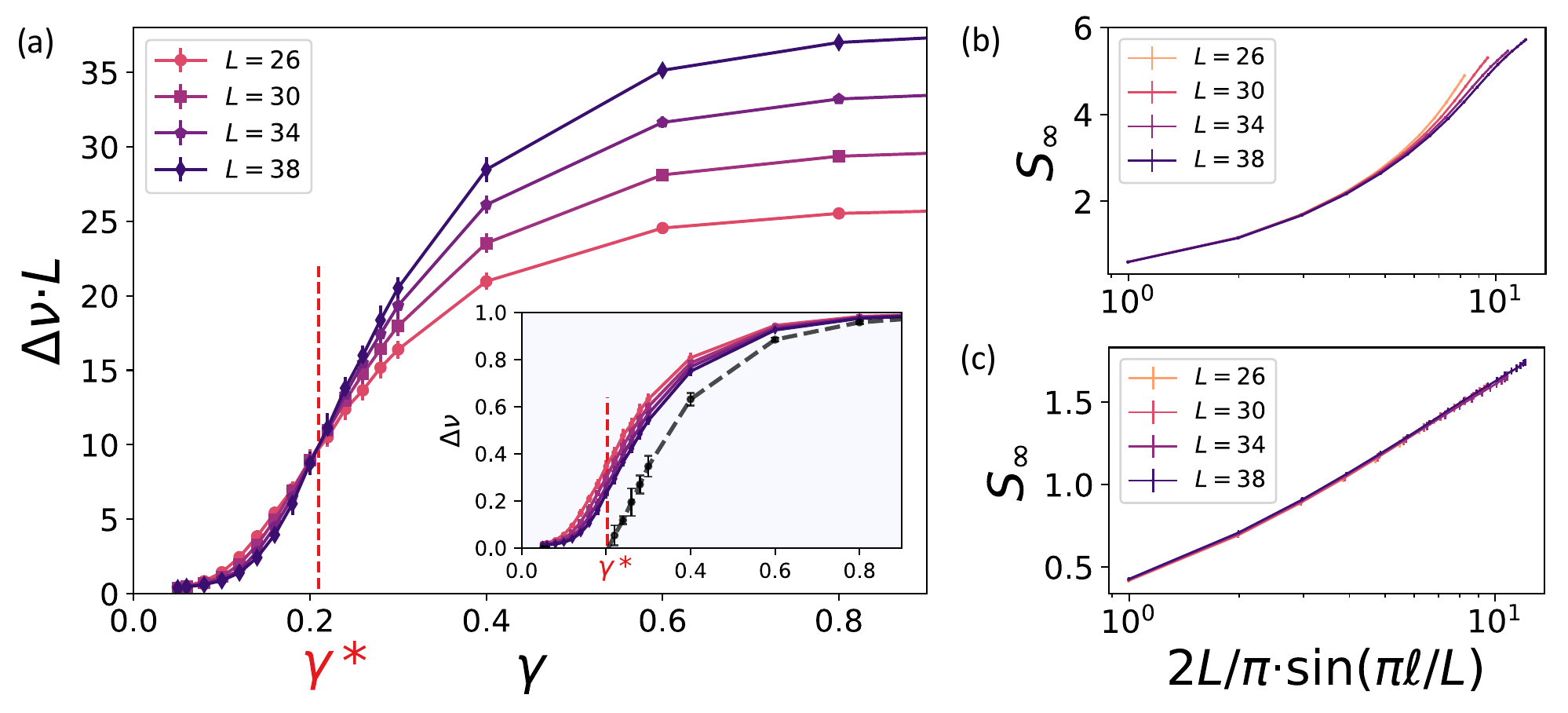}
\caption{(a) Finite-size scaling of the slopes at the opening of the gap for the interacting case with measurements of $n_j$. The insert shows the scaling of the gaps $\Delta\nu = \nu_{N}-\nu_{N-1}$. 
Increasing the size $L$ adds new eigenvalues $\nu_\alpha\in[0,1]$, producing an obvious scaling of the gaps. Rescaling all spectra in the same domain $\alpha=0,...,L-1\to\alpha/L\in[0,1]$, the slopes $\Delta\nu L$ grow unbounded if a gap remains in the thermodynamic limit. We expect a closure of the gap if $\Delta\nu$ decreses faster than the average spacing $\sim1/L$, i.e. if the slope tends to $0$. The crossing signals the opening of the gap at the critical point $\gamma=0.21$, where the entanglement entropies collapse on the CFT curve (c). The dashed line is a linear extrapolation in $1/L$ of the gap. For comparison, we plot in (b) the scaling of the entropies in the volume-law for $\gamma=0.05$.} 
\label{fig: gaps U}
\end{figure}

\subsection{MIPT and single-particle occupation spectra}

The total non-Gaussianity highlights the similarities between non-unitary “interactions” provided by measurements and the actual interactions, but it is not informative about the MIPT. We obtain new insights on the transition by looking directly at the eigenvalues $\nu_\alpha$. 
Consider first the interacting Hamiltonian with $U=1$ and $n_j$ measurements. 
%
%
Deep in the Zeno phase, the large fraction of measured sites keeps the state close to a collection of eigenstates of $n_j$ and the state is almost single-particle. Correspondingly, half of the natural orbitals are almost fully occupied $\nu_\alpha\approx1$ while the other half is almost unoccupied $\nu_\alpha\approx0$ as shown in Fig. \ref{fig: main}. There is a jump $\Delta\nu = \nu_{N}-\nu_{N+1}\to$ for $\gamma\to\infty$. Decreasing $\gamma$, the discontinuity remains at all rates until $\gamma^*=0.21$: for $\gamma\le\gamma^*$ the spectrum appears smooth and $\Delta\nu$ closes in the thermodynamic limit, as shown in Fig. \ref{fig: gaps U}. 
A continuous spectrum is expected for an ergodic phase as a consequence of ETH \cite{Bera2017}, and the entanglement at $\gamma^*$ reproduces the scale-invariant behaviour ${S}_\infty(\ell) = \alpha\log\left(\frac{2L}{\pi}\sin\frac{\pi\ell}{L}\right)+s_0$, where the chord length in the argument of the $\log$ takes into account the finite size of the system \cite{Calabrese2004}. Increasing the interaction strength $U$ pushes to higher values the volume-law and the discontinuity moves correspondingly: the appearence of the area law is associated to the opening of a gap in the spectrum of the one-body reduced density matrix, which is then an indicator of the measurement induced phase transition and we identify $\gamma^*=\gamma_c$. Similarities between MIPTs in d spatial dimension and localized systems in d+1 have been noted before and our work enlarges the list, since an analogous phenomenon has been described in the context of many-body localization (MBL) \cite{Poboiko2023,Fava2023,Chahine2023,Znidaric2008,Pal2010,Abanin2019}. In the case of MBL, the role of the long-time steady state is played by high-energy eigenstates of the Hamiltonian and the pinning of states is introduced by a disordered potential instead of stochastic measurements, but again the ergodic phase is analogously characterized by a smooth spectrum with a gap that opens up at the transition \cite{Bera2015,Bera2017,Hopjan2021}.

\begin{figure}[t]
\centering
    \includegraphics[width=0.48\textwidth]{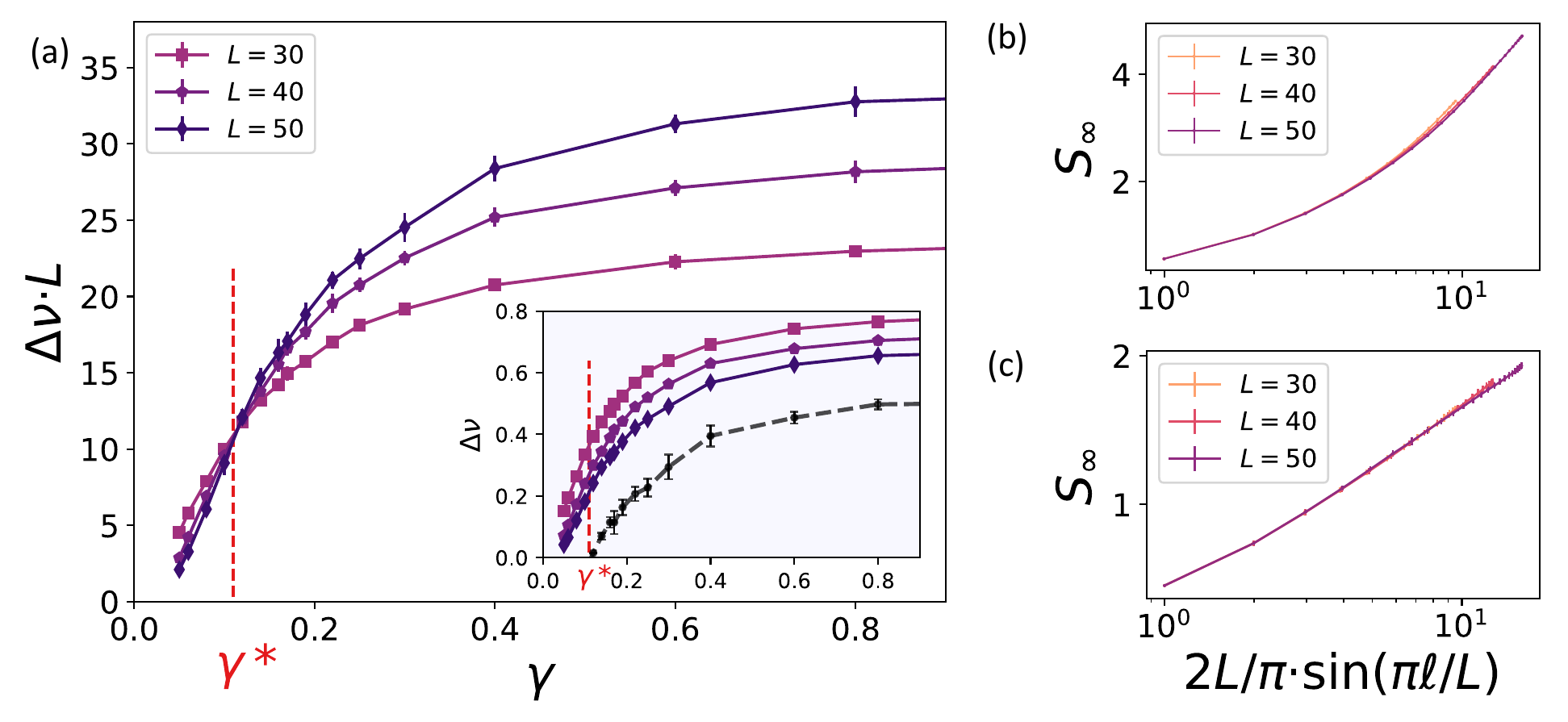}
\caption{(a) Finite-size scaling of the gaps and the associated slopes for the free Hamiltonian with current measurements. We see again a crossing at $\gamma=0.11$, where the entanglement has a clear critical behaviour (c), indicating the opening of the gap and the transition towards the area law. At $\gamma=0.05$ we see tiny deviations from the critical collapse that can be interpreted as a signal of the finite-size crossover.} 
\label{fig: gaps J}
\end{figure}
For current measurements (Fig. \ref{fig: gaps J}), in the area law region the spectrum of the correlation matrix is gapped as expected, but it never approaches the step function $\nu_\alpha=1$ for $\alpha<L/2$, $\nu_\alpha=0$ for $\alpha>L/2$ even in the $\gamma\to\infty$ limit. 
This is again a consequence of the non-commuting nature of currents 
%
The gap remains present in the thermodynamic limit until $\gamma^*=0.11$, where it displays critical behaviour. We estimate $\gamma_c=\gamma^*$ as the critical rate at which the area law emerges, and for $\gamma<\gamma_c$ the system is in a distinct dynamical phase. 
%
%
With non-Gaussian measurements, the single-particle gap supports the presence of a stable measurement-induced phase transition in a system of free fermions and the sub-extensive regime is the finite-size crossover associated to a volume-law phase. The lower $NG$ shown in Fig.\ref{fig: NG} explains why the volume-law appears less clearly with current measurements at the rates observed, in a similar way as how reducing the coupling constant $U$ lowers the total $NG$ and correspondingly pushes the transition towards smaller values of $\gamma$. For small measurement rates the effects of the non-commutativity of the currents are of higher order in $\gamma dt$ and negligible, therefore this transition is unrelated to the measurement frustration mechanism \cite{Ippoliti2021} and only due to the non-Gaussian nature of quantum trajectories.

\section{Conclusions}

In this work, we have discussed the role of non-Gaussianity in measurement induced phase transitions by studying the quantum trajectories generated by current measurements. Non-Gaussian states are characterized by a non-trivial spectrum of their one-body reduced density matrix and we observe an opening of a gap in correspondence of the transition towards the area law. Our findings show that there are no fundamental differences between introducing non-Gaussianity unitarily through interactions and non-unitarily through measurements as far as the entanglement transition is concerned, as in the thermodynamic limit both stabilize the MIPT. However, a clear volume-law growth of the entanglement has not been observed. The reason lies in the measurement protocol itself, that has broken Gaussianity but not strongly enough, in a similar way to an interaction term with a small coupling that would push the transition towards the rare-measurement regime. 

\section{Acknowledgements}

We thank Alessandro Romito and Henning Schomerus for the insightful suggestions. Tensor network calculations wew performed with the aid of the TeNPy library \cite{tenpy}. This work has been supported by the ERC under grant agreement n.101053159 (RAVE), by a Google Quantum Research Award, by the PNRR MUR project PE0000023-NQSTI, and by the PRIN 2022 (2022R35ZBF) - PE2 - ``ManyQLowD''. E. T. was partly supported by the MIUR Programme FARE (MEPH), by QUANTERA DYNAMITE PCI2022-132919, and by the EU-Flagship programme Pasquans2.

\appendix
\section{Conditional and unconditional dynamics}
\label{app: Conditional and unconditional dynamics}

Here we present a derivation of the stochastic Schrödinger equation given in the main text. Consider for simplicity measurements of a single local observable $\mathcal{Q}_j$ at a fixed site $j$ with only two outcomes $q_{1},\,q_{2}$ and associated projectors $\Pi_{q_1},\,\Pi_{q_2}$ s.t.  $\sum_q \Pi_q = \mathbb{I}$. Starting from an initial pure state $|\psi\rangle$, a projective measurement of $\mathcal{Q}$ is made with probability $\gamma dt$, which will result in one of the two outcomes with the proper probability $p_\psi(q) = \langle\psi|\Pi_q|\psi\rangle$. Otherwise, we just let the state evolve unitarily. This quantum map is encoded in the Kraus operators
$K_0 = \mathbb{I} - iHdt - \frac{K}{2},\,$ $K_1 = \sqrt{\gamma dt} \Pi_{q_1},\, K_2 = \sqrt{\gamma dt}\, \Pi_{q_2},$
where the non-unitary part $K=K_1^\dagger K_1 + K_2^\dagger K_2 = {\gamma dt}$ in the no-click term is a re-normalization  fixed by the conservation of probability $\sum_q K_q^{\dagger}K_q = \mathbb{I}$.
The Hamiltonian evolution could be included in the other cases as well, but it would result in a difference of higher order ${o}(dt)$ and we neglect it. Then 
\begin{equation}
|\psi(t)\rangle\,\,\to\,\,|\psi(t+dt)\rangle = \frac{K_m|\psi(t)\rangle}{\sqrt{\langle K_m^{\dagger}K_m \rangle_t}}
\end{equation}

\noindent with probability $\langle\psi(t) | K_m^{\dagger}K_m |\psi(t)\rangle\equiv\langle K_m^{\dagger}K_m \rangle_t$
takes into account both the classical probability of making a measurement and the quantum probability generated by Born's rule. We obtain the stochastic update rule 
\begin{align*}
    |\psi(t)\rangle\,&\to \,\frac{\Pi_{q_1}|\psi(t)\rangle}{\sqrt{\langle\Pi_{q_1}\rangle_t}}\,\,,\,\,\, \text{ prob.}=\gamma dt\,p_t(q_1)\,,\\
    &\to\, \frac{\Pi_{q_2}|\psi(t)\rangle}{\sqrt{\langle\Pi_{q_2}\rangle_t}}\,\,,\,\,\, \text{prob.}=\gamma dt\,p_t(q_2)\,,\\
    &\to\, \frac{[\,\mathbb{I}-iHdt-K/2]|\psi(t)\rangle}{\sqrt{1-\langle K \rangle_t}}\,\,,\,\, \text{prob.}=1-\gamma dt\,.    
\end{align*}
Monitoring a complete observable provides $K=\gamma dt\propto\mathbb{I}$, which simplifies the evolution without measurements to $|\psi(t+dt)\rangle=(\mathbb{I}-iHdt)|\psi(t)\rangle$ up to $O(dt)$. We want to follow the dynamics of a pure state conditioned on the measurement results, generating random trajectories of pure states called quantum trajectories. The trajectories can be characterized by defining the stochastic variables $N_q(t)$, which count the number of measurements with outcome $q$ on a given realization until time $t$. Their increments satisfy $dN_q(t)=0,1\,\Rightarrow\,\,dN_q(t)^2=dN_q(t)$. On average, the increment equals the probability of making a measurement with the corresponding outcome $\overline{dN_q}(t) = \langle K_q^{\dagger}K_q \rangle_t  = \gamma dt\,p_t(q)$ and it defines a Poisson process. Using $(1+x)^\alpha = 1+\alpha x + O(x^2)$, we can take advantage of $dN_q$ to write the evolution compactly as
\begin{widetext}
\begin{equation}
    |\psi(t+dt)\rangle = |\psi(t)\rangle + 
    \sum_q dN_q(t) \frac{\Pi_q|\psi\rangle}{\sqrt{\langle\Pi_q\rangle_t}} + \left(1-\sum_q dN_q(t)\right)\left[\mathbb{I}-iHdt-\frac{K}{2}+\frac{\langle K \rangle_t}{2}\right]|\psi(t)\rangle\,.
\end{equation}
Recalling that $K$ is of order $dt$, $dN_q$ can be neglected in the last term apart from product with the identity. Then
\begin{equation}
    d|\psi(t)\rangle  =  \Biggl\{ \sum_{q} dN_{q}(t) \Biggl(\frac{\Pi_q}{\sqrt{\langle \Pi_q \rangle_t}} -1 \Biggr)
    - dt\Biggl[iH + 
        \frac{\gamma}{2}\sum_{q}\Bigl(\Pi_q-\langle \Pi_q\rangle_t\Bigr) \Biggr] \Biggr\}\,|\psi(t)\rangle\,,
\end{equation}
\end{widetext}
which is the stochastic Schrödinger equation presented in the main text, besides the trivial generalization to take into account measurements on all sites $j$. Here we focused on the case of strong projective measurements, but an analogous procedure could be followed to obtain the quantum jump SSE for a generic POVM. The unconditional dynamics, instead, physically corresponds to a monitoring process where the measurement outcomes are discarded and after each measurement the state is averaged over them. In this way a single measurement is enough to turn the initial pure state into a mixture
\begin{equation}
\begin{split}
    \rho(t+dt) &= \overline{|\psi(t+dt)\rangle\langle\psi(t+dt)|}\\
    &= \sum_{m}\,prob(m) \,\rho_m = \sum_m\,K_{m}\rho(t)K_m^{\dagger}\,,
\end{split}
\end{equation}
where $\rho_m=K_m\rho(t)K_m^{\dagger}$ is the resulting density matrix conditioned on the outcome. This defines the Kraus representation of a Markovian quantum channel, which evolves in time according to the Lindblad equation
\begin{equation}
    \frac{d\rho}{dt} = -i[H,\rho]+\gamma\sum_{q}\Bigl(\Pi_q\rho \Pi_q-\frac{1}{2}\{\Pi_q,\rho\}\Bigr)\,.
\end{equation}
For example, monitoring $\sigma^z_j$ on a chain of $L$ spins (which is equivalent to measuring $n_j$ on a fermionic chain under Jordan-Wigner) corresponds to the substitution $\Pi_q\to\Pi_j^\pm = \frac{1}{2}(\mathbb{I}\pm\sigma^z_j)\,,\,\,\sum_q\to\sum_{j,\pm}$. Then
\begin{equation}
    \frac{d\rho}{dt} = -i[H,\rho]+\frac{\gamma}{2}\sum_{j}\Bigl(\sigma^z_j\rho \sigma^z_j-\frac{1}{2}\{\sigma^z_j,\rho\}\Bigr)\,,
\end{equation}
and the measurements have provided a dephasing noise term
which drives the system towards infinite temperature. The same is typically true for all these master equations, because measurements provide hermitian Lindblad operators which always make the Liouvillian unital. Unital means that $\dot{\rho}=\mathcal{L}(\rho)=0$ for $\rho\propto\mathbb{I}$, and the maximally mixed state is expected to be the stationary solution towards which the mean state relaxes.

\section{Total non-Gaussianity}
\label{app: Total non-Gaussianity}

Let $|\psi\rangle$ be a pure fermionic Gaussian state, i.e. a Slater determinant, and consider its correlation matrix $C_{ij}(\psi) = \langle c_i^{\dagger}c_j \rangle$. Then $|\psi\rangle = \prod_\alpha (a_\alpha^{\dagger})^{\nu_\alpha} |vac\rangle$ on some set of single-particle states defined by the creation operator $a_\alpha^\dagger$ and the correlation matrix is diagonal on the basis of states $a_\alpha^\dagger|vac\rangle$ with eigenvalues $\nu_\alpha=0,1$. In the main text, we have described how to characterize the departure of pure a state from Gaussianity by inspecting the spectrum of its one-body reduced density operator $\rho^{(1)}(\psi) = \sum_{ij} C_{ij}(\psi)\,c^\dagger_i c_j$ . However, this criterion does holds only for pure states. In general it remains possible to quantify how far a mixed state is distant from being Gaussian (that for mixed states means an exponential of a quadratic form of creation/annihilation operators). A resource theory of non-Gaussianity was put forward in \cite{Genoni2008,Genoni2010} for bosonic states, here we present a direct extension for fermions. The distance of two quantum states $\rho,\sigma$ is often described in terms of their relative entropy
\begin{equation}
    S(\rho||\sigma) = \text{tr}[\rho(\log\rho-\log\sigma)]\,.
\end{equation}
$S(\rho||\sigma)$ is not an actual metric (nor a quasi-metric) because it does not respect symmetry and triangle inequality, but it is a meania<ngful measure of distinguishability since the probability of not distinguishing $\sigma$ from $\rho$ after $N$ measurements on $\sigma$ is $\exp\{-NS(\rho||\sigma)\}$ \cite{Vedral2002}. Since Gaussian states are completely characterized by their correlation matrix, given a non-Gaussian state $\rho$ and $C_{ij}=\text{tr}(\rho\, c^\dagger_i c_j)$ we can construct its Gaussian partner $\rho_G$. As it is natural to expect, $\rho_G$ is actually the closest Gaussian state to $\rho$ since $\min_\sigma \{S(\rho||\sigma)\} = S(\rho||\rho_G)$ when $\sigma$ is varied among the set of Gaussian states \cite{Marian2013}. Then we can quantify the amount of non-Gaussianity in a state by the relative entropy $NG(\rho)=S(\rho||\rho_G)$. In the normal mode basis Gaussian states can be expressed as tensor products of exponentials of quadratic single-mode Hamiltonians, then $\log \rho_G$ is a second degree polynomial operator in the ladder operators and the calculation of $\text{tr}\left[\rho\log\rho_G\right]$ involves only combinations of second moments $\langle c^\dagger_i c_j\rangle=\text{tr}[\rho\, c^\dagger_i c_j]=\text{tr}[\rho_G\, c^\dagger_i c_j]$. As a consequence, $\rho$ can be replaced by $\rho_G$ and the total non-Gaussianity reduces to 
\begin{equation}
\label{eq: non-Gaussianity}
    NG(\rho)\equiv S(\rho||\rho_G) = S(\rho_G)-S(\rho)\,.
\end{equation}
For a pure state $|\psi\rangle$, $S(\rho)\equiv S(\psi)=0$ and $NG(\psi)$ reduces to the entropy calculated as if it were Gaussian
\begin{equation}
\label{eq: Gaussian entropy}
    NG(\psi) = - \sum_\alpha \bigl[\nu_\alpha\log\nu_\alpha+ (1-\nu_\alpha)\log{(1-\nu_\alpha)}\bigr]\,.
\end{equation}
The properties of the relative entropy ensure that: 
\begin{enumerate}[label=(\roman*)]
    \item $NG(\rho)\ge0$ ($=0$ iff $\rho=\rho_G$)
    \item $NG(\rho_1\otimes\rho_2) = NG(\rho_1)+NG(\rho_2)$
    \item $NG(U\rho U^\dagger) = NG(\rho)$ for any $U=e^{-iH}$ where $H=H^\dagger$ is a quadratic hamiltonian
    \item $NG(\text{tr}_A\rho)\le NG(\rho)$ under a partial trace over an arbitrary subsystem $A$
    \item $NG(\mathcal{G}(\rho))\le NG(\rho)$ for any Gaussian quantum channel $\mathcal{G}$\,.
\end{enumerate} 
The property (i) is just a re-stating of Klein's inequality and it means that Gaussian states are indeed free states. En passant, notice that (i) is also an alternative proof of the fact that Gaussian states maximize the Von Neumann entropy at a fixed correlation matrix. (ii) and (iii) grant that appending free states and unitary Gaussian transformations are free operations. The proof of (ii) is trivial, while (iii) can be proven by noting that $ \text{tr}(U\rho U^\dagger \, c_i^\dagger c_j) = \text{tr}(\rho\, U^\dagger  c_i^\dagger c_j U)$ and $(U\rho U^\dagger)_G$ can be constructed by calculating the correlation matrix on the transformed operators $c_i\to U^\dagger c_i U$. For an infinitesimal $U=e^{-i\lambda H}\approx \mathbb{I}-i\lambda H$ $c_i$ evolves according to the Heisenberg equation $\dot{c}_i(\lambda) = i[H,c_i].$ Assuming particle number conservation, we can take $H = \sum_{j,k}\,h_{jk}c_j^\dagger c_k$ and $\dot{c}_i(\lambda) = -i \sum_j\,h_{ij}c_j$. If $H$ is quadratic $[H,c_i]$ contains only linear terms in $c_i$ (or $c_i^\dagger$ if the conservation law is broken) and the evolved $c_i$ remains a single particle operator. The finite transformation $c_i \to U^\dagger c_i U$ is equivalent to a unitary rotation in the single-particle $c_i$ space $\vec{c} \to \mathcal{U} \vec{c}\,$, $\mathcal{U} = e^{-ih}$ and the correlation matrix changes only up to a similarity transformation. For pure states this is already sufficient to prove that NG is invariant, since the two matrices have the same spectrum. In general $(U\rho U^\dagger)_G = U\rho_G U^\dagger$ because they have the same correlation matrix, then $NG(U\rho U^\dagger) = S(U\rho U^\dagger || U\rho_GU^\dagger) = S(\rho || \rho_G) = NG(\rho)$ as Von Neumann entropies are invariant under global unitaries.
The quantum relative entropy is is monotonously decreasing under partial traces and this property is inherited by $NG$: $(\text{tr}_A\rho)_G = \text{tr}_A(\rho_G)$ implies $NG(\text{tr}_A\rho) \le NG(\rho)$. From a resource-theory point of view, this means that discarding a part of the system is correctly regarded as a free operation.
A Gaussian channel $\mathcal{G}$ is a CPTP map that evolves Gaussian states into other Gaussian states, therefore it can be expressed as $\mathcal{G}(\rho) = \text{tr}_E\left[U(\rho\otimes\sigma_G)U^\dagger\right]$,
where the system $\rho$ is coupled to a Gaussian environment $\sigma_G$, that is traced out affter a common evolution with a Gaussian unitary. Putting the statements above together with (ii) the property (v) follows: $NG(\mathcal{G}(\rho)) \,\le \, NG(U(\rho\otimes\sigma_G)U^\dagger) = NG(\rho) + NG(\sigma_G) = NG(\rho)\,.$
Notice the importance of having a Gaussian environment: if the state of the environment is generic we only have $NG(\mathcal{G}(\rho))\le NG(\rho)+NG(\sigma)$.
For pure states $NG(\psi)$ given by Eq. \eqref{eq: Gaussian entropy} is concave. Expanding the channel in its Kraus representation implies the stronger inequality $\sum_q\,p_q NG(\mathcal{G}_q\psi)\le NG(\psi)$, which means that non-Gaussianity cannot decrease in average also along quantum trajectories, when the average is performed after the calculation of the non-linear quantity.

\section{Details on the simulation}
\label{app: Details on the simulation}

When the dynamics is restricted to special classes of quantum states, such as Gaussian or Clifford states, the system can be simulated in an exact and efficient way on a classical computer. Here we are interested in the dynamics of generic interacting quantum system and we simulate our models using tensor networks, which provide an approximation of the desided state by truncating its quantum correlations. Area-law entangled states on a 1D spin chain can be naturally encoded as a matrix product state (MPS), while volume-law states can be well represented as an MPS if the bond dimension, the parameter which controls the entanglement truncation, is chosen appropriately. To represent fermionic degrees of freedom we need to map our operators in terms of spin variables, keeping track of Jordan-Wigner strings.
Having defined $\sigma^\pm = \frac{1}{2}(\sigma^x\pm i \sigma^y)$ and assuming $|n=0\rangle \equiv |\uparrow\,\rangle$\, $|n=1\rangle \equiv |\downarrow\,\rangle$, the Jordan-Wigner mapping is
\begin{equation}
    c_j = \prod_{k=1}^{j-1}\sigma^z_j\cdot \sigma^+_j\,\,,\,\,\,\,c_j^\dagger = \prod_{k=1}^{j-1}\sigma^z_j\cdot \sigma^-_j\,\,,\,\,\,\,n_j =\frac{1-\sigma^z_j}{2}\,.
\end{equation}
After the transformation, the Hamiltonian \eqref{eq: hamiltonian} becomes
\begin{equation}
\begin{split}
    H = &-\frac{1}{4}\sum_{j=1}^L \left(\sigma^x_i\sigma^x_{i+1}+\sigma^y_i\sigma^y_{i+1}\right) \\
    &+ \frac{U}{4}\sum_{j=1}^L\left(1-\sigma^z_j-\sigma^z_{j+1}+\sigma^z_j\sigma^z_{j+1}\right)\,.
\end{split}
\end{equation}
The local relation $\sigma^z_j = 1-2n_j$ has two consequences: projective measurement of the local occupations are equivalent to measurements of the spin and the $U(1)$ symmetry of the original model corresponds exactly to the $U(1)$ symmetry of the spin chain generated by the conserved magnetization $M = \sum_j\sigma^z_j$. The correspondence therefore extends to the current operator
\begin{equation}
    J_j = -\frac{1}{4}\left(\sigma^y_j\sigma^x_{j+1}-\sigma^x_j\sigma^y_{j+1}\right)\,,
\end{equation}
which is the opposite of the usual spin current because according to $\sigma^z_j=1-2n_j$ the magnetization increases in the direction where the occupations decrease. Local occupations are associated the projectors
\begin{equation}
    \Pi_j^{n=0,1} = \frac{\mathbb{I}\pm\sigma_j^z}{2}\,.
\end{equation}
The current has an eigenvalue with double degeneracy $J_j=0$ and two non-degenerate eigenvalues $J_j=\pm 1/2$, associated to the eigenstates:
\begin{gather}
\label{eq: eigenstates}
    |\psi_{0,0}\rangle = |\!\uparrow\uparrow\,\rangle\,,\,\,\,|\psi_{0,1}\rangle = |\!\downarrow\downarrow\,\rangle\,,\\
    |\psi_{\pm}\rangle = \frac{1}{\sqrt{2}} \bigl(|\!\uparrow\downarrow\,\rangle\mp i|\!\downarrow\uparrow\,\rangle) \,.
\end{gather}
Measuring $J_j$ corresponds to applying the projectors 
\begin{gather}
    \Pi_j^{J=0} = \frac{\mathbb{I}+\sigma^z_j\sigma^z_{j+1}}{2} \,,\\
    \Pi_j^{J=\pm1/2} = \frac{\mathbb{I}-\sigma^z_j\sigma^z_{j+1}\mp\sigma^y_j\sigma^x_{j+1}\pm\sigma^x_j\sigma^y_{j+1}}{4}\,.
\end{gather} 
The quantities that we study in order to characterize the dynamics are the entanglement entropy of connected subystems and the fermionic correlation matrix. Some care is needed when calculating the entanglement entropy of fermionic modes through the Jordan-Wigner transformation, because it is a non-local mapping: the Hilbert spaces of the whole fermionic and spin chains are in correspondence, but not those of subsystems. However, for a connected subchain $A$ all Jordan-Wigner strings connecting $i,j\in A$ are fully contained in $A$, so no problems arise and $S_A$ can be calculated directly in terms of the dual spin variables. Assuming $i<j$, the correlation matrix is
\begin{equation}
    \langle c_i^\dagger c_j\rangle = \langle \sigma_i^-\prod_{k=i}^{j-1}\sigma_k^z\,\sigma^+_j \rangle
\end{equation}
Along a quantum trajectory the state remains pure. We represent it as an MPS and 
its dynamics can be simulated with a modified TEBD algorithm. We divide it into $N$ discrete time steps of width $dt$, each composed of a unitary part $\exp(-iHdt)|\psi\rangle$ and measurements. The unitary evolution is implemented with the usual TEBD, with an even-odd Trotter decomposition scheme. Then, if the measurement rate is $\gamma$, we extract a random number $x_j$ for each site (or bond) $j$ and if $x_j<\gamma dt$ a projective measurement of the observable $Q_j$ is performed. In that case, we compute the probabilities $p_\psi(q)=\langle \psi|\Pi_j^{q}|\psi\rangle$ for each eigenstate $q$ and according to them we extract which projector $\Pi_j^{q}$ we apply at the measured $j$.  This completes a step and the procedure is iterated $N$ times. In the case of current measurements it is important to apply the projectors in a randomized order, because it may happen that many operators are measured at the same time step and since they do not commute the order is relevant. Measurements act on specific sites, breaking the translation and reflection invariance of the dynamics for single trajectories. The random location of measurements restores the symmetry on the mean state and the average entanglement entropy is a function only of subsystem size $\overline{S}_A=\overline{S}(|A|)$. For projectors applied in a fixed order, e.g. from left to right, the mean state loses the symmetry as well and for $A=\{1,...,\ell\}$ if $\overline{S}(\ell)\neq\overline{S}(L-\ell)$, even if $\overline{S}_A=\overline{S}_{A^C}$. Randomizing the position of measurements is preferrable because this asymmetry is fictitious and disappears in the continuum limit, since for $dt\to0$ the probability of having neighbouring measurements at the same time step is ${O}(dt^2)$.

\nocite{*}
\bibliography{apssamp} 



\end{document}